\documentclass[a4paper,english,prb,aps,twocolumn,floatfix,showpacs,tightenlines,superscriptaddress,amsmath,amssymb]{revtex4}
\pdfoutput=1
\usepackage[T1]{fontenc}
\usepackage[latin9]{inputenc}
\usepackage{verbatim}
\usepackage{amsmath}
\usepackage{graphicx}
\usepackage{amssymb}

\makeatletter

\newcommand{\1}{\leavevmode{\rm 1\ifmmode\mkern  -4.8mu\else\kern -.3em\fi I}}

\usepackage{babel}
\makeatother

\begin{document}

\title{The fidelity approach to the Hubbard model}

\author{L. Campos Venuti}

\affiliation{ISI Foundation for Scientific Interchange, Villa Gualino, Viale Settimio
Severo 65, I-10133 Torino, Italy}

\author{M. Cozzini}

\affiliation{ISI Foundation for Scientific Interchange, Villa Gualino, Viale Settimio
Severo 65, I-10133 Torino, Italy}

\affiliation{Dipartimento di Fisica, Politecnico di Torino, Corso Duca degli Abruzzi
24, I-10129 Torino, Italy}

\author{P. Buonsante}

\affiliation{CNISM, Unit� di Ricerca Torino Politecnico, Corso Duca degli Abruzzi
24, I-10129 Torino, Italy}

\affiliation{Dipartimento di Fisica, Politecnico di Torino, Corso Duca degli Abruzzi
24, I-10129 Torino, Italy}

\author{F. Massel}

\affiliation{Dipartimento di Fisica, Politecnico di Torino, Corso Duca degli Abruzzi
24, I-10129 Torino, Italy}

\author{N. Bray-Ali}

\affiliation{Department of Physics and Astronomy, University of Southern California
Los Angeles, California 90089-0484, USA}

\author{P. Zanardi}

\affiliation{Department of Physics and Astronomy, University of Southern California
Los Angeles, California 90089-0484, USA}

\affiliation{ISI Foundation for Scientific Interchange, Villa Gualino, Viale Settimio
Severo 65, I-10133 Torino, Italy}


\begin{abstract}
We use the fidelity approach to quantum critical points to study the
zero temperature phase diagram of the one-dimensional Hubbard model.
Using a variety of analytical and numerical techniques, we analyze
the fidelity metric in various regions of the phase diagram, with
particular care to the critical points. Specifically we show that
close to the Mott transition, taking place at on-site repulsion $U=0$
and electron density $n=1$, the fidelity metric satisfies an hyper-scaling
form which we calculate. This implies that in general, as one approaches
the critical point $U=0,\, n=1$, the fidelity metric tends to a limit
which depends on the path of approach. At half filling, the fidelity
metric is expected to diverge as $U^{-4}$ when $U$ is sent to zero.
\end{abstract}

\date{\today}

\maketitle

\section{Introduction}

Recently a novel characterization of phase transitions has been advocated
\citep{PZ-PK06}. This is the so called fidelity approach to critical
phenomena \citep{ZB,ZCG07,CGZ07,CIZ07,YLG,GKNL,zhou,buonsante07,hamma-lidar07,ZZL,chen07,ZZWL,zhu07,yang07,ZCG07PRL}
that relies solely on the state of the system and does not require
the knowledge of the model Hamiltonian and its symmetry breaking mechanism.
Two states of the system at nearby points in parameter space are compared
by computing their overlap (the fidelity). Since quantum phase transitions
are major changes in the structure of the ground state, it is natural
to expect that, when one crosses a transition point the fidelity will
drop abruptly. To make the analysis more quantitative one considers
the second derivative of the fidelity with respect to the displacement
in parameter space. Remarkably this second derivative, more in general
the Hessian matrix, defines a metric tensor (the fidelity metric hereafter)
in the space of pure states \citep{ZCG07PRL}. A super-extensive scaling
of the fidelity metric corresponds to the intuitive idea of a fidelity
drop. Indeed, it was shown in \citep{lcv07} that, at regular points
the fidelity metric scales extensively with the system size, and a
super-extensive behavior implies criticality. However the converse
is not true in general; in order to observe a divergence in the fidelity
metric a sufficiently relevant perturbation (in the renormalization
group sense) is needed \citep{lcv07}. Loosely speaking the more relevant
the operator the stronger the divergence of the fidelity metric. The
Berezinskii-Kosterlitz-Thouless (BKT) transition is peculiar in this
respect as it is driven by a marginally relevant perturbation, i.e.~with
the smallest possible scaling dimension capable of driving a transition.
This gives rise to an infinite order transition and as such the BKT
does not rigorously fit in the simple scaling argument given in \citep{lcv07}. 

Surprisingly, contrary to the na\"ive expectation, Yang has shown
\citep{yang07} that in the particular instance of BKT transition
provided by the XXZ model, the fidelity metric diverges algebraically
as a function of the anisotropy. This is an appealing feature since
observing a singularity at a BKT transition is generally a difficult
task%
\footnote{An almost standard route is that of adding an electric/magnetic field
and to measure the corresponding stiffness which experiences a jump
at the transition (see i.e.~\citep{laf01})%
}.

In this paper we analyzed, with a variety of analytical and numerical
techniques, the one dimensional Hubbard model primarily aiming at
assessing the power and limitations of the fidelity approach for infinite
order QPT ($n=1,\, U\rightarrow0$). We believe it is useful to list
here the main accomplishment of our analysis: i) An exact calculation,
on the free gas line $U=0$, shows that the fidelity metric $g$ presents
a cusp at half filling and a $1/n$ divergence at low density $n$
respectively. ii) Using Bosonization techniques, we observe a divergence
of the form $g\sim n^{-2}$ in the regime $U\ll n$ when $n\rightarrow0$.
iii) Resorting to Bethe Ansatz we are able to interpolate between
the regime where the Luttinger liquid parameter $K$ approaches the
BKT value $1/2$ and that where $K\to1$, which describes the free-Dirac
point. We show that the fidelity metric satisfies an hyper-scaling
equation which can also be extended to finite sizes. iv) We calculate
the hyper-scaling function in the thermodynamic limit by solving Bethe-Ansatz
integral equations while at half filling by resorting to exact diagonalization.
As a consequence, when approaching the transition point $U=0,\, n=1$
the fidelity metric tends to a limit which depends on the path of
approach. On the particular path $U\to0,\, n=1$, an algebraic divergence
of the form $g\sim U^{-4}$ is expected, on the basis of numerical
results. 

In the 1D Hubbard model the BKT transition %
\footnote{The term BKT is justified in the sense that the underlying effective
theory is the sine-Gordon model (see later). In the condensed matter
literature the term Mott transition is more commonly used. %
} occurs exactly at half filling as soon as the on site interaction
$U$ is switched on, inducing a gap in the charge excitation spectrum.
Away from half filling instead all modes are gapless for any $U$
and the system is a Luttinger liquid. Since at half filling, the only
gapless point is at $U=0$, the kind of BKT transition offered by
the Hubbard model is different from that featured by the XXZ model.
In that case one continuously arrives at the transition point from
a gapless phase by tuning the anisotropy parameter. This difference,
in turn, makes more difficult the analysis of the fidelity metric
in the Hubbard model.

\section{Preliminaries}

The one dimensional Hubbard model is given by\begin{equation}
H=-t\sum_{i,\sigma}\left(c_{i,\sigma}^{\dagger}c_{i+1,\sigma}+\mathrm{h.c.}\right)+U\sum_{i}n_{i,\uparrow}n_{i,\downarrow}-\mu\sum_{i}n_{i}\,,\label{eq:ham-hubbard}\end{equation}
where $n_{i,\sigma}=c_{i,\sigma}^{\dagger}c_{i,\sigma}$ , $n_{i}=n_{i,\uparrow}+n_{i,\downarrow}$
and we will be concerned with the repulsive / free gas case when $U\ge0$.
Due to the symmetries of the model \citep{HubbBook} it is sufficient
to limit the analysis to filling smaller or equal to one half i.e.~$n\equiv\langle n_{i}\rangle\le1$.
It is well known \citep{solyom79,voit95} that for $n<1$ the model
is in the Luttinger liquid universality class for any value of the
interaction $U$. Spin and charge degrees of freedom separate and
their respective excitations travel at distinct velocities $v_{s}$
and $v_{c}$. Both charge and spin modes are therefore gapless. Exactly
at half filling ($n=1$) the system becomes an insulator and develops
a charge gap $\Delta E_{c}=\mu_{+}-\mu_{-}=E\left(N+1\right)-2E\left(N\right)+E\left(N-1\right)$,
where $E\left(N\right)$ is the ground state energy with $N$ particles.
The gap opens up exponentially slow from $U=0$, and the point $U=0,\, n=1$
is a transition of BKT type \citep{itzykson}. The length scale $\xi=2v_{c}/\Delta E_{c}$
describes the size of soliton-antisoliton pairs, in the insulator.
As we approach the critical point at $U=0,\, n=1$, these pairs unbind
and proliferate, allowing the system to conduct. 

In the fidelity approach one is interested in the overlap between
ground states at neighboring points of the coupling constants (say
a vector $\lambda$): $F\left(\lambda\right)=\left|\langle\psi\left(\lambda\right)|\psi\left(\lambda+d\lambda\right)\rangle\right|$.
Remarkably, the second order term in the expansion of the fidelity
defines a metric in the space of (pure) states:\[
F\left(\lambda\right)=1-\frac{1}{2}G_{\mu,\nu}d\lambda^{\mu}d\lambda^{\nu}+O\left(d\lambda^{3}\right)\,,\]
where\[
G_{\mu,\nu}=\mathrm{Re}\left[\langle\partial_{\mu}\psi_{0}|\partial_{\nu}\psi_{0}\rangle-\langle\partial_{\mu}\psi_{0}|\psi_{0}\rangle\langle\psi_{0}|\partial_{\nu}\psi_{0}\rangle\right],\]
and $\psi_{0}=\psi\left(\lambda\right)$ and $\partial_{\mu}=\partial/\partial\lambda^{\mu}$.
Actually, at regular points $\lambda$ of the phase diagram, $G_{\mu,\nu}$
is an extensive quantity \citep{lcv07} so that it is useful to define
the related intensive metric tensor $g_{\mu,\nu}\equiv G_{\mu,\nu}/L$.
With reference to Hamiltonian (\ref{eq:ham-hubbard}) it is natural
to investigate the behavior of the fidelity under variations of the
interaction parameter $U$. The possibility of analyzing variations
of the chemical potential, though appealing does not fit in the fidelity
approach as the ground states at different $\mu$ belong to different
super-selection sectors. Hence now on we will solely be interested
in $g_{U,U}$ and we will simply write $g$ in place of $g_{U,U}$.
In Ref.~\citep{ZCG07PRL} it was shown that it can be written in
the following form\begin{equation}
g=\frac{1}{L}\sum_{n>0}\frac{\left|V_{n,0}\right|^{2}}{\left(E_{n}-E_{0}\right)^{2}},\quad V=\sum_{i}n_{i,\uparrow}n_{i,\downarrow}\label{eq:g}\end{equation}
where $E_{n},\,|n\rangle$ are the eigen-energies and corresponding
eigenstates of the Hamiltonian \eqref{eq:ham-hubbard} (with repulsion
$U$ and filling $n$), $|0\rangle$ corresponds then to the ground
state and $V_{i,j}=\langle i|V|j\rangle$. In passing, we would like
to notice that despite the apparent similarity between Eq.~\eqref{eq:g}
and the second derivative of the energy $E''\left(\lambda\right)$
(a similarity stressed in Ref.~\citep{chen08}), it is possible to
show -- using the Rayleigh-Schr\"odinger series -- that the metric
tensor is in fact related to the third (and first) energy derivative,
via\[
G=\frac{1}{V_{0,0}}\sum_{i,j>0}\frac{V_{0,i}V_{i,j}V_{j,0}}{\left(E_{i}-E_{0}\right)\left(E_{j}-E_{0}\right)}-\frac{E'''}{E'}.\]
Moreover, in cases where $E'''\left(\lambda\right)$ is bounded in
the thermodynamic limit, one obtains the interesting kind of factorization
relation $\langle0|V|0\rangle\langle0|V\mathcal{G}\left(E_{0}\right)^{2}V|0\rangle=\langle0|V\mathcal{G}\left(E_{0}\right)V\mathcal{G}\left(E_{0}\right)V|0\rangle$
valid in the thermodynamic limit, where $\mathcal{G}$ is the resolvent
$\mathcal{G}\left(E\right)$=$\left(\1-|0\rangle\langle0|\right)\left(H-E\right)^{-1}\left(\1-|0\rangle\langle0|\right)$. 

In the rest of the paper we will be concerned with the analysis of
the metric tensor with special care to the BKT transition point. The
phase diagram of the Hubbard model is depicted in Fig.~\ref{fig:phase_diagram}.
The model has been solved by Bethe Ansatz in Ref.~\citep{lieb_wu68}.
We will tackle the problem using a variety of techniques. On the free-gas
$U=0$ line, an explicit calculation is possible at all fillings.
Around the region $U=0$ and filling away from $n=0$ and $n=1$ bosonization
results apply. Instead, close to the points $U=0$ and $n=1$ we will
cross results from bosonization with Bethe-Ansatz in order to extend
bosonization results up to the transition points. We will show that
the behavior of the metric is encoded in a scaling function. Away
from half filling the scaling function is computed integrating Bethe-Ansatz
equation, while at half filling by resorting to exact diagonalization.

\begin{center}
\begin{figure}[th]
\begin{centering}
\includegraphics[width=70mm,keepaspectratio]{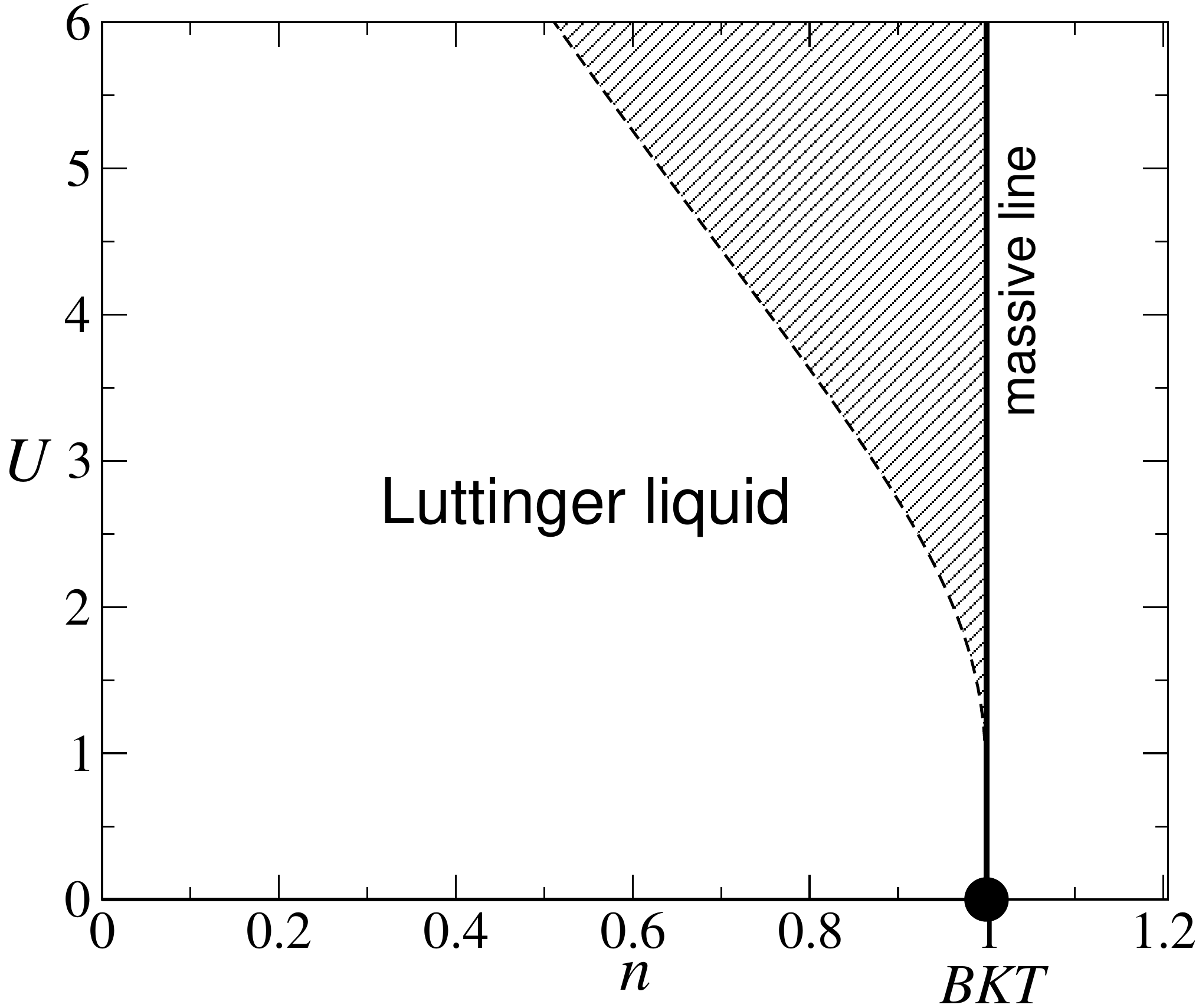}
\par\end{centering}

\caption{Phase diagram of the repulsive Hubbard model. The region $0<n<1$
is in the Luttinger liquid (LL) universality class. The hatched area
corresponds to the condition $\delta\xi\left(U\right)<1$, where $\xi\left(U\right)$
is the correlation length at half filling. Approaching the critical
point $\left(n=1,U=0\right)$ from within (the complementary of) this
region the LL parameter $K_{c}$ approaches $1/2$, the BKT value
($1$, the free-Dirac value). \label{fig:phase_diagram}}

\end{figure}

\par\end{center}

\section{Exact analysis at $U=0$}

At $U=0$ the complete set of eigenfunctions of Hamiltonian (\ref{eq:ham-hubbard})
is given by a filled Fermi sea and particle-hole excitations above
it. It is then possible to apply directly Eq.~(\ref{eq:g}).

\subsection{Half filling\label{sub:Half-filling}}

We first treat the half filled case $n=1$, where the Fermi momentum
lies at $k_{F}=\pi/2$. Writing the interaction in Fourier space as
$V=L^{-1}\sum_{k,k',q}c_{k'-q,\downarrow}^{\dagger}c_{k+q,\uparrow}^{\dagger}c_{k,\uparrow}c_{k',\downarrow},$and
going to the thermodynamic limit, Eq.~(\ref{eq:g}) becomes\begin{equation}
g=\frac{1}{\left(2\pi\right)^{3}}\int_{\left[-\pi,\pi\right]^{3}}\frac{n_{k}\left(1-n_{k+q}\right)n_{k'}\left(1-n_{k'-q}\right)}{\left(\epsilon_{k+q}-\epsilon_{k}+\epsilon_{k'-q}-\epsilon_{k'}\right)^{2}}dkdk'dq,\label{eq:start}\end{equation}
where $\epsilon_{k}=-2t\cos\left(k\right)$ is the $U=0$ single particle
dispersion and $n_{k}=\vartheta\left(-\epsilon_{k}\right)$ ($=n_{k,\uparrow}=n_{k,\downarrow}$
in absence of magnetic field) are the fermionic, zero-temperature,
filling factors. Using $\epsilon\left(k,q\right)\equiv\epsilon_{k+q}-\epsilon_{k}=4\sin\left(q/2\right)\sin\left(k+q/2\right)$
and substituting $k'\rightarrow-k'$ we obtain\begin{multline*}
g=\frac{1}{\left(2\pi\right)^{3}}\int_{0}^{\pi}dq\frac{1}{8\sin\left(q/2\right)^{2}}\int_{-\pi}^{\pi}dk\int_{-\pi}^{\pi}dk'\\
\times\frac{n\left(k,q\right)n\left(k',q\right)}{\left[\sin\left(k+q/2\right)+\sin\left(k'+q/2\right)\right]^{2}}.\end{multline*}
Changing variables to $p=k+q/2$, $p'=k'+q/2$ and making a shift
of $\pi/2$ we obtain finally\begin{equation}
g=\frac{1}{\left(2\pi\right)^{3}}\int_{0}^{\pi}dq\frac{J\left(q\right)}{8\sin\left(q/2\right)^{2}}=\frac{1}{24\pi^{2}}=0.00422\label{eq:gn1}\end{equation}
where we defined\begin{eqnarray*}
J\left(q\right) & = & 4\int_{0}^{q/2}dp\int_{0}^{q/2}dp'\frac{1}{\left[\cos\left(p\right)+\cos\left(p'\right)\right]^{2}}\\
 & = & -2+8\frac{\ln\left(\cos\left(q/2\right)\right)}{\cos\left(q\right)-1},\end{eqnarray*}
and correctly $J\left(q\right)=J\left(-q\right)>0$. 

A related interesting issue is that of the finite size scaling of
the metric tensor $g$ i.e.~the way in which $g_{L}$ at length $L$
converges to its thermodynamic value (see also \citep{YLG}). In Ref.~\citep{lcv07}
it was shown that in a gapless regime scaling analysis predicts $g_{L}\sim L^{-\Delta_{g}}$
apart from regular contributions which scale extensively (and contribute
to $g$ with a constant). Here $\Delta_{g}=2\Delta_{V}-2\zeta-1$
where $\Delta_{V}$ is the scaling dimension of $V$ in the renormalization
group sense and $\zeta$ is the dynamical critical exponent. On the
line $U=0$ one has $\Delta_{V}=2$ as $V$ is a product of two independent
free fields, while $\zeta=1$ when $n\neq0$ due to the linear dispersion
of excitations at low momenta. This implies that at leading order
$g_{L}\sim A+BL^{-1}$. One should however be careful that logarithmic
corrections are not captured by the scaling analysis of Ref.~\citep{lcv07}
and they might be present due to the BKT transition occurring at this
point. Let us try to clarify this issue. Looking at Eq.~\eqref{eq:gn1},
as a first approximation, we might think that the finite size $g_{L}$
is well approximated by the Riemann sum of $F\left(q\right)\equiv J\left(q\right)/\sin\left(q/2\right)^{2}$:\[
g_{L}\simeq\frac{S_{L}}{\left(2\pi\right)^{3}8},\quad S_{L}=\frac{2\pi}{L}\sum_{n=1}^{L/2}F\left(\frac{2\pi}{L}n\right).\]
Now $F\left(q\right)$ diverges logarithmically around $\pi^{-}$:
$F\left(q\right)=-4\ln\left(\frac{\pi-q}{2}\right)+O\left(\left(\pi-q\right)^{2}\right)$,
and since the Riemann sum of the logarithm converges to its integral
as $\left(A+B\ln L\right)/L+O\left(L^{-2}\right)$ we would conclude\begin{equation}
g_{L}-\frac{1}{24\pi^{2}}=\frac{A+B\ln L}{L}+\frac{C}{L^{2}}+O\left(L^{-4}\right).\label{eq:g_FSS}\end{equation}
However, a detailed analysis of the finite size $g_{L}$ reveals that
a \emph{cancellation} occurs between two logarithmic corrections so
that actually $B=0$ in Eq.~(\ref{eq:g_FSS}). The exact finite size
$g_{L}$ is composed of two terms. One term is a triple sum which,
in the thermodynamic limit, corresponds to the triple integral in
Eq.~(\ref{eq:start}). The other term is a double sum which originates
from zero transferred momentum contribution and vanishes when $L\rightarrow\infty$.
We numerically verified that both terms contain a $\ln L/L$ part
when $L\rightarrow\infty$, but their contribution is equal and opposite
so as to cancel out exactly from $g_{L}$. The absence of logarithmic
corrections can clearly be seen in the inset of Fig.~\ref{fig:g_vs_density}
where the finite size $g_{L}$ is plotted against $1/L$.

\subsection{Away from half filling}

Similar considerations can be done away from half-filling. Eq.~(\ref{eq:start})
still holds, simply in this case $k_{F}\neq\pi/2$. We assume $k_{F}=n\pi/2<\pi/2$
(anyway a particle hole transformation implies $g\left(n\right)=g\left(2-n\right)$).
First note that the integral over $q$ can be recast as $2\int_{0}^{\pi}dq$.
The filling factors constrain the momenta to $\left|k\right|<k_{F}$
and $\left|k+q\right|>k_{F}$. If $q<2k_{F}$ this implies $k_{F}-q<k<k_{F}$.
Instead if $q>2k_{F}$ the sum over $k$ is unconstrained: $-k_{F}<k<k_{F}$.
Thus we obtain\begin{multline*}
g=\frac{2}{\left(2\pi\right)^{3}}\left\{ \int_{0}^{2k_{F}}dq\int_{k_{F}-q}^{k_{F}}dk\int_{k_{F}-q}^{k_{F}}dk'\right.\\
\left.+\int_{2k_{F}}^{\pi}dq\int_{{-k}_{F}}^{k_{F}}dk\int_{-k_{F}}^{k_{F}}dk'\right\} \\
\times\frac{1}{16\sin\left(q/2\right)^{2}\left[\sin\left(k+q/2\right)+\sin\left(k'+q/2\right)\right]^{2}}.\end{multline*}
Changing variables as before and defining \[
J\left(a,b\right)\equiv\int_{a}^{b}dp\int_{a}^{b}dp'\frac{1}{\left[\sin\left(p\right)+\sin\left(p'\right)\right]^{2}},\]
we obtain \begin{multline*}
g=\frac{1}{64\pi^{3}}\left\{ \int_{0}^{2k_{F}}dq\frac{J\left(k_{F}-q/2,k_{F}+q/2\right)}{\sin\left(q/2\right)^{2}}\right.\\
\left.+\int_{2k_{F}}^{\pi}dq\frac{J\left({-k}_{F}+q/2,k_{F}+q/2\right)}{\sin\left(q/2\right)^{2}}\right\} .\end{multline*}
In Fig.~\ref{fig:g_vs_density} one can see a plot of $g\left(n,U=0\right)$
as a function of the total density $n=N_{\mathrm{tot}}/L=2k_{F}/\pi$.
It is possible to show that in the very dilute regime $n\rightarrow0$,
the fidelity metric $g$ diverges in a simple algebraic way \[
g\left(n\rightarrow0,U=0\right)\sim\frac{1}{n}.\]
This divergence can also be simply understood by resorting to the
scaling arguments reported in \citep{lcv07}. There it was shown that,
in the thermodynamic limit $g\sim\left|\mu-\mu_{c}\right|^{\Delta_{g}/\Delta_{\mu}}$
where now $\Delta_{\mu}$ is the scaling dimension of the field $\mu$.
On the line $U=0$ as already noticed $\Delta_{V}=2$ while now $\zeta=2$
as $n\rightarrow0$ to account for the parabolic dispersion. The chemical
potential scaling exponent is $\Delta_{\mu}=2$ in the dilute Fermi
gas \citep{sachdevQPT}. All in all we obtain $g\sim\left|\mu-\mu_{c}\right|^{-1/2}\sim n^{-1}$
since $n\sim\left|\mu-\mu_{c}\right|^{1/2}$ which agrees with the
explicit calculation. 

The finite size scaling of $g_{L}$ for different filling $0<n<1$
is the same as that observed at $n=1$ and is dictated by Eq.~\eqref{eq:g_FSS}
with $B=0$, as can bee seen in the inset of Fig.~\ref{fig:g_vs_density}.

Finally note that, since $g\left(n,U=0\right)$ is symmetric around
$n=1$, $g\left(n\right)$ has a local maximum at that point with
a cusp. The origin of this discontinuity is not well understood at
the moment but reveals a signature of the transition occurring at
this point. 

\begin{center}
\begin{figure}[th]
\begin{centering}
\includegraphics[width=78mm,keepaspectratio]{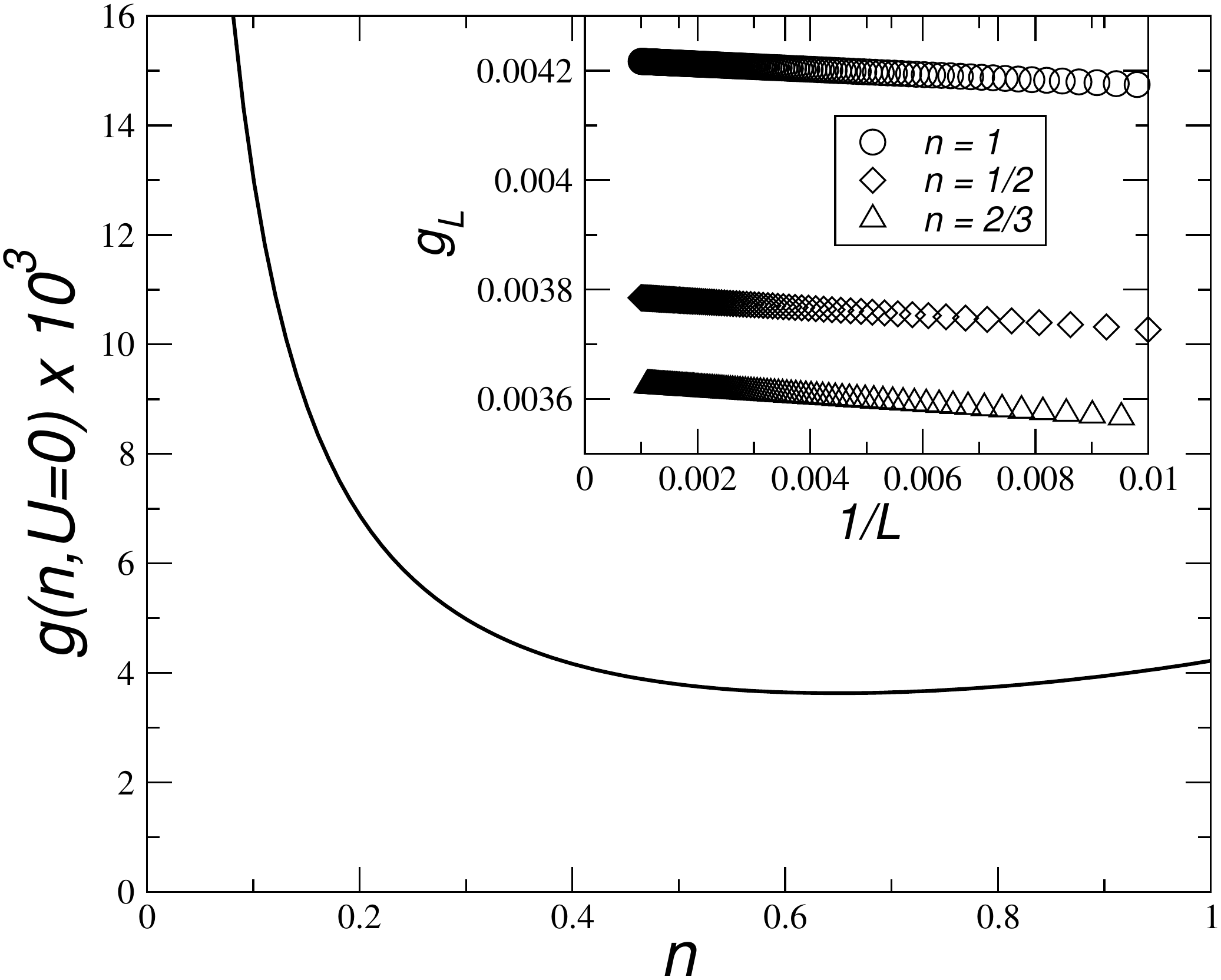}
\par\end{centering}

\caption{Fidelity metric $g$ as a function of the total density at $U=0$.
The singularity at $n\rightarrow0$ is of the form $n^{-1}$. In the
inset the finite size scaling of $g_{L}$ for some different fillings
is shown. The approach to the thermodynamic value is given by Eq.~\eqref{eq:g_FSS}
with $B=0$. In fact fitting the data points with Eq.~\eqref{eq:g_FSS}
gives values of $B/A$ of the order of $10^{-4}$ and a chi-square
of the same order as the one obtained with $B=0$.\label{fig:g_vs_density}}

\end{figure}

\par\end{center}

\section{Bosonization approach}

It is well known \citep{solyom79,emery79,HubbBook} that for $U\ge0$
and away from half filling ($n=1$), the low energy, large distance
behavior of the Hubbard model, up to irrelevant operators, is described
by the Hamiltonian\begin{eqnarray}
H & = & H_{s}+H_{c}\label{eq:ham-eff}\\
H_{\nu} & = & \frac{v_{\nu}}{2}\int d^{2}x\left[K_{\nu}\Pi_{\nu}\left(x\right)^{2}+\frac{1}{K_{\nu}}\left(\partial_{x}\Phi_{\nu}\right)^{2}\right],\,\,\nu=s,\, c\,.\nonumber \end{eqnarray}
Charge and spin degrees of freedom factorize and are described respectively
by $H_{c}$ and $H_{s}$. The Luttinger liquid parameters $K_{c,s}$
are related to the long distance, algebraic decay of correlation functions,
while $v_{c,s}$ are the speed of elementary (gapless) charge and
spin excitations. From bosonization, and setting the lattice constant
$a=1$, one finds, for small $U$, \begin{equation}
K_{c}=1-\frac{U}{{2\pi v}_{F}}+\cdots,\label{eq:K-bosoniz}\end{equation}
where the Fermi velocity is $v_{F}=2t\sin\left(k_{F}\right)$ and
$k_{F}=\pi n/2$. Instead the Luttinger parameter $K_{s}$ is fixed
to $K_{s}=1$ due to spin rotation invariance. Exactly at $n=1$ there
appears another term (an Umklapp term) in the charge sector which
is marginally relevant and is responsible for the opening of a mass
gap. In this case the effective theory is the sine-Gordon model. 

Since the fidelity of two independent theories factorizes the metric
tensor $g$ is additive and we obtain $g=g_{s}+g_{c}$. In Ref.~\citep{yang07}
the fidelity metric of a free boson theory has been calculated to
be given by \begin{equation}
g_{\nu}=\frac{1}{8}\left(\frac{1}{K_{\nu}}\frac{dK_{\nu}}{dU}\right)^{2}.\label{eq:g-free-bosons}\end{equation}
In our case $g_{s}=0$ as $K_{s}$ does not vary so that $g=g_{c}+g_{s}=g_{c}$.
Using Eq.~(\ref{eq:K-bosoniz}) one obtains a formula valid up to
zeroth order in $U$: \begin{equation}
g=\frac{1}{2\left(4\pi v_{F}\right)^{2}}+O\left(U\right).\label{eq:g-bosoniz}\end{equation}
Some comments are in order. The expansion \eqref{eq:K-bosoniz} is
actually an expansion around $U=0$ valid when $U\ll v_{F}$. When
we move toward the BKT critical point one has $v_{F}\rightarrow2$
and $g\rightarrow1/\left(128\pi^{2}\right)$. This value is $3/16$
the number calculated directly at $U=0$ in Sec.~\ref{sub:Half-filling}.
We believe that this discrepancy is due to lattice corrections which
are neglected in formula \eqref{eq:g-free-bosons}. Approaching the
low density critical point $U=0,\, n=0$ Eq.~\eqref{eq:g-bosoniz}
predicts that, in a narrow region $U\ll n$, the fidelity metric diverges
as $g\sim n^{-2}$. This contrasts with the result $g\sim n^{-1}$
obtained at $U=0$, as one would expect since $U$ is a relevant perturbation.
In fact, in the diluted regime, the low-energy effective theory is
that of a spinful non-relativistic gas with delta interactions \citep{HubbBook}.
There one still has $n\sim\left|\mu-\mu_{c}\right|^{1/2}$ and dynamical
exponent $\zeta=2$. Thus if we take $\Delta_{\mu}=1$, then using
the conventional scaling analysis we would find $g\sim n^{-2}$, consistent
with the bosonization result.

\section{Hyper-scaling of fidelity metric near the Metal-Insulator critical
point\label{sec:Luttinger-parameter}}

The bosonization expression Eq.~\eqref{eq:K-bosoniz} is an expansion
of $K_{c}\left(n,U\right)$ around $U=0$ where $K_{c}$ reaches its
free Dirac value 1. In the whole stripe $U\ge0$, $0\le n\le1$, $K_{c}\left(n,U\right)$
is a bounded function ranging between $1/2$ and $1$ \citep{schluz90}.
The maximal value $K_{c}=1$ is obtained in the segment $U=0$. Instead
the minimal value $K_{c}=1/2$ is attained at the lines $n=0$ and
$n=1$ and in the strong coupling limit, i.e.~$K_{c}\rightarrow1/2$
for $U\gg\left|t\right|$. This considerations show that, from Eq.~\eqref{eq:g-free-bosons},
$g$ can be infinite only at the points $U=0$ and $n=0,\,\mbox{or}\,1$
where $K_{c}$ is discontinuous. In particular we are interested to
the vicinity of the transition point $U=0,\, n=1$ which we will call
simply (with some abuse) BKT point. Calling $\xi\left(U\right)$ the
correlation length at half filling, and $\delta=1-n$ the doping concentration,
it can be shown (see later) that the Luttinger liquid parameter $K_{c}$
tends to $1/2$ when the BKT point is approached from the region $\delta\xi\left(U\right)\ll1$.
Instead $K_{c}\to1$ when the BKT point is approached from $\delta\xi\left(U\right)\gg1$.
Given this discontinuity of $K_{c}$ it seems difficult to interpolate
between the two regimes $\delta\xi\left(U\right)\ll1$ and $\delta\xi\left(U\right)\gg1$.
However we will show that such an interpolation is indeed possible
and that the fidelity metric $g$ satisfies an hyper-scaling relation
valid when $\delta\xi$ ranges over many order of magnitudes (see
figure \ref{fig:g_scaling}). 

In Refs.~\citep{kawakami90,frahm90} an efficient characterization
of the Luttinger liquid parameter $K_{c}$ in terms of Bethe Ansatz
results has been found. $K_{c}$ is related to the so called dressed
charge function $Z_{k}$ through \begin{equation}
K_{c}=Z_{Q}^{2}/2,\label{eq:K_dressedcharge}\end{equation}
where the wave vector $Q$ is a generalization of the Fermi wave-vector
in the interacting regime, and has to be determined by Bethe-Ansatz
equations. We will now argue that, in the metallic phase, the dressed
charge function $Z_{Q}$ satisfies the following scaling relation:\begin{equation}
Z_{Q}\left(U,\delta\right)=\Phi_{Z}\left(\xi\left(U\right)\delta\right).\label{eq:Z_scaling}\end{equation}
Here $\xi\left(U\right)$ is the correlation length at half filling
($\delta=0$) defined via $\xi=v_{c}/\Delta E_{c}$ where $v_{c}$
is the charge carries (holons) velocity and $\Delta E_{c}$ is the
(charge) gap, and $\Phi_{Z}$ is a scaling function. Obviously a similar
scaling relation holds for the Luttinger parameter $K_{c}$ through
Eq.~\eqref{eq:K_dressedcharge}. The scaling relation holds as long
as the interaction is not too strong, say $U\lesssim1$. A similar
scaling relation has been conjectured in Ref.~\citep{stafford93}
for the charge stiffness. 

Using equation \eqref{eq:g-free-bosons} we obtain the following scaling
relation for the fidelity metric of the Hubbard model in the metallic
phase close to the BKT point\begin{equation}
g\left(U,\delta\right)=\left(\frac{d\ln\xi}{dU}\right)^{2}\Phi_{g}\left(\xi\delta\right),\label{eq:g_scaling}\end{equation}
where we introduced the scaling function $\Phi_{g}\left(x\right)=\left(1/2\right)\left(x\Phi_{Z}'\left(x\right)/\Phi_{Z}\left(x\right)\right)^{2}$.
Following Ref.~\citep{stafford93} it is natural to conjecture a
more general hyper-scaling relation for the fidelity metric valid
also at finite size. Building the other dimensionless quantity with
the correlation length $\xi$ and the size $L$, we obtain\begin{equation}
g\left(U,\delta,L\right)=\left(\frac{d\ln\xi}{dU}\right)^{2}Y\left(\xi\delta,\xi/L\right).\label{eq:g_hyper-scaling}\end{equation}

\begin{center}
\begin{figure}[th]
\begin{centering}
\includegraphics[width=78mm]{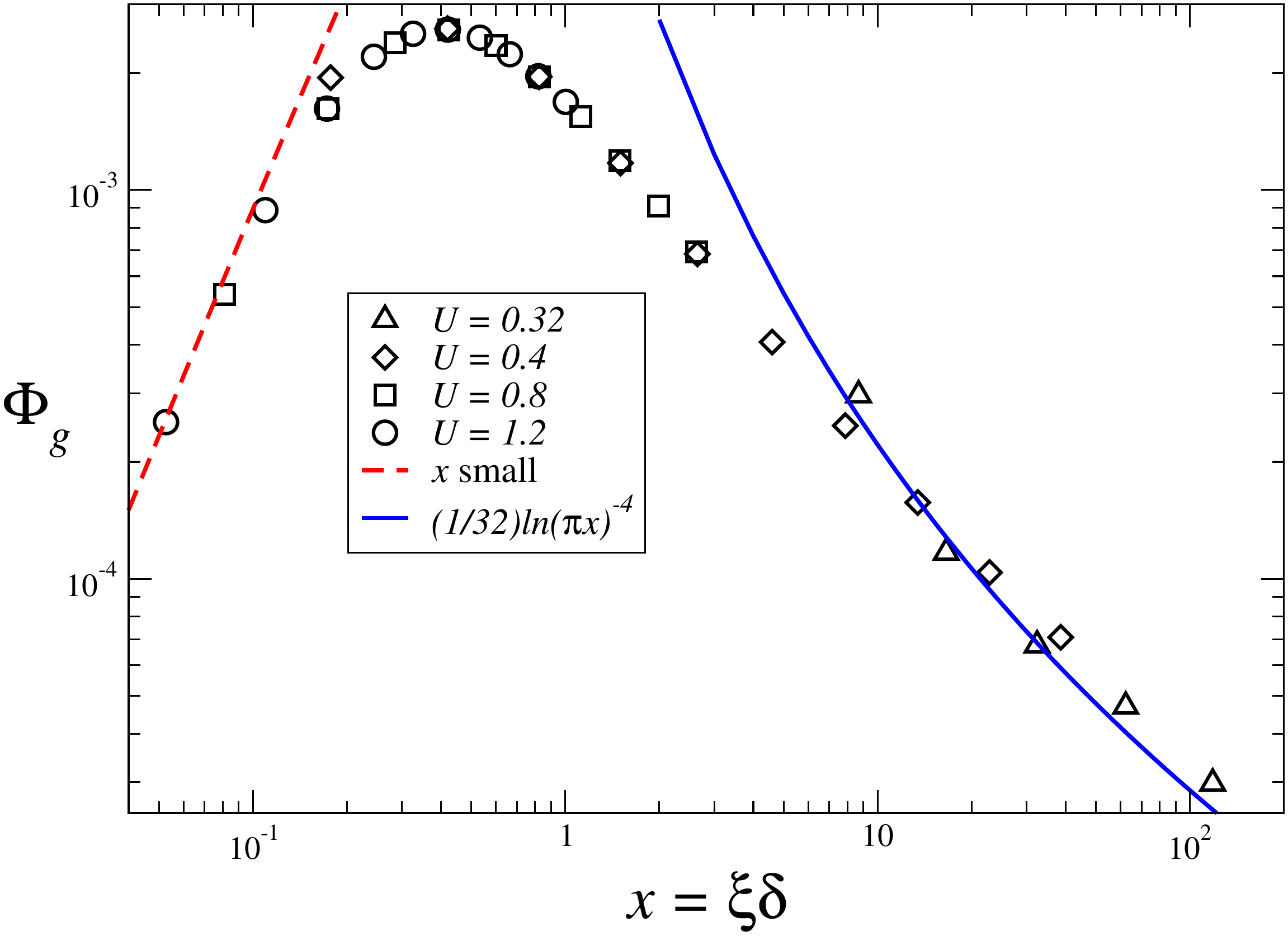}
\par\end{centering}

\caption{(color online). Scaling function $\Phi_{g}$ for the fidelity metric
as a function of the scaling parameter $\xi\delta$. Data are obtained
by solving numerically the integral equations for the dressed charge
at different interaction strength $U$. The dashed line is obtained
expanding the Bethe-Ansatz equations for the Luttinger parameter in
the regime $\xi\delta\ll1$ up to second order in $\delta$. The solid
line results from integrating the RG BKT equations and is valid when
$\xi\delta\gg1$. \label{fig:g_scaling}}

\end{figure}
 
\par\end{center}

We will now verify the hyper-scaling relation Eq.~\eqref{eq:g_hyper-scaling}
away from half-filling in the thermodynamic limit ($\xi/L=0$) using
various analytical techniques and at finite size resorting to exact,
Lanczos, diagonalization.

\subsection{Away from half filling}

The scaling relation Eq.~(\ref{eq:g_scaling}) (and implicitly Eq.\eqref{eq:Z_scaling})
can be verified analytically in the two limits $x=\xi\delta\to0$
and $x\to\infty$. We present the results in terms of the Luttinger
parameter $K_{c}$ which has more physical relevance. In Refs.~\citep{frahm90,schad91},
the Bethe Ansatz equations for the dressed charge have been solved
around $\delta=0$. At leading order in $\delta$, they found for
the Luttinger parameter\begin{equation}
K_{c}=\frac{1}{2}+a\left(U\right)\delta+\frac{1}{2}\left[a\left(U\right)\delta\right]^{2}+O\left(\delta^{3}\right).\label{eq:kappac}\end{equation}
The function $a\left(U\right)$ is studied in the appendix and is
approximately given by the following expansion, for $U/2\pi\ll1$\[
a\left(U\right)\approx\frac{\ln\left(2\right)}{\sqrt{U}}e^{2\pi/U}.\]
Not surprisingly, this has the same form as the soliton length, $\xi\left(U\right)$.
In fact, for the regime of interest, $U/2\pi\lesssim1$, $v_{c}\to2$
and $\xi\left(U\right)=\pi a\left(U\right)/2\ln\left(2\right)$ \citep{stafford93}.
This implies that in the region $x\ll1$, the metric scaling function
behaves, at leading order, as \[
\Phi_{g}\left(x\right)=\left(\frac{\ln4}{\sqrt{2}\pi}x\right)^{2}+O\left(x^{3}\right).\]

Conversely, in the opposite regime $\xi\delta\gg1$ (i.e.~$U\to0$,
$\delta$ small) integrating the renormalization-group BKT equations
\citep{kolomeisky96}, one is able to improve the bosonization result
Eq.~\eqref{eq:K-bosoniz} and we obtain \begin{eqnarray*}
K_{c} & = & 1-\frac{U/\left(4\pi\right)}{1-U/\left(2\pi\right)\ln\left(1/\pi\delta\right)}\\
 & = & 1-\frac{1}{2}\left[\ln\left(\pi\xi\delta\right)\right]^{-1}+\ldots.\end{eqnarray*}
Accordingly, the scaling function $\Phi_{g}\left(x\right)$ has the
following asymptotic behavior when $x\to\infty$\[
\Phi_{g}\left(x\right)\simeq\frac{1}{32}\left[\ln\left(\pi x\right)\right]^{-4}.\]
In the limit $U\to0$ we recover bosonization's result $g\to1/128\pi^{2}$. 

To verify the scaling relation Eq.~\eqref{eq:g_scaling} also in
the intermediate regime $\xi\delta\approx1$, we solved numerically
the Bethe-Ansatz equations for the dressed charge \citep{HubbBook}.
To obtain the dressed charge function $Z_{k}$ we need also the density
of wave numbers $\rho_{k}$. They are solutions of the following integral
equations\begin{align}
\rho_{k} & =1-\cos\left(k\right)\!\!\int_{-Q}^{Q}\!\! dq\cos qR\left(\sin k-\sin q\right)\rho_{q}\label{eq:Z-int}\\
Z_{k} & =1+\int_{-Q}^{Q}\!\! dq\cos qR\left(\sin k-\sin q\right)Z_{q}\label{eq:rho-int}\end{align}
with $R\left(x\right)$ given by\[
R\left(x\right)=\frac{1}{2\pi}\int_{-\infty}^{\infty}d\omega\frac{e^{i\omega x}}{1+e^{U\left|\omega\right|/2}}.\]
The wave-vector $Q$ is determined by fixing the electronic density
$n=\int_{-Q}^{Q}\rho_{k}dk$. Integrating numerically Eqs.~\eqref{eq:Z-int}
and \eqref{eq:rho-int} for different value of the coupling strength
and doping fraction, we are able to verify the scaling relation Eq.~\eqref{eq:g_scaling}
over many order of magnitudes. The result is plotted in figure \ref{fig:g_scaling}. 

We would like to point out that, since close to the BKT point the
relevant variable is $\xi\left(U\right)\delta$, the limit of the
fidelity metric when $U\to0,\, n\to1$ depends on the path of approach.
However, as we have shown, the combination $g\left(d\ln\xi/dU\right)^{-2}$is
a perfectly well defined function of $\xi\delta$. 

{}

\begin{figure}[b]
\includegraphics[width=8.5cm]{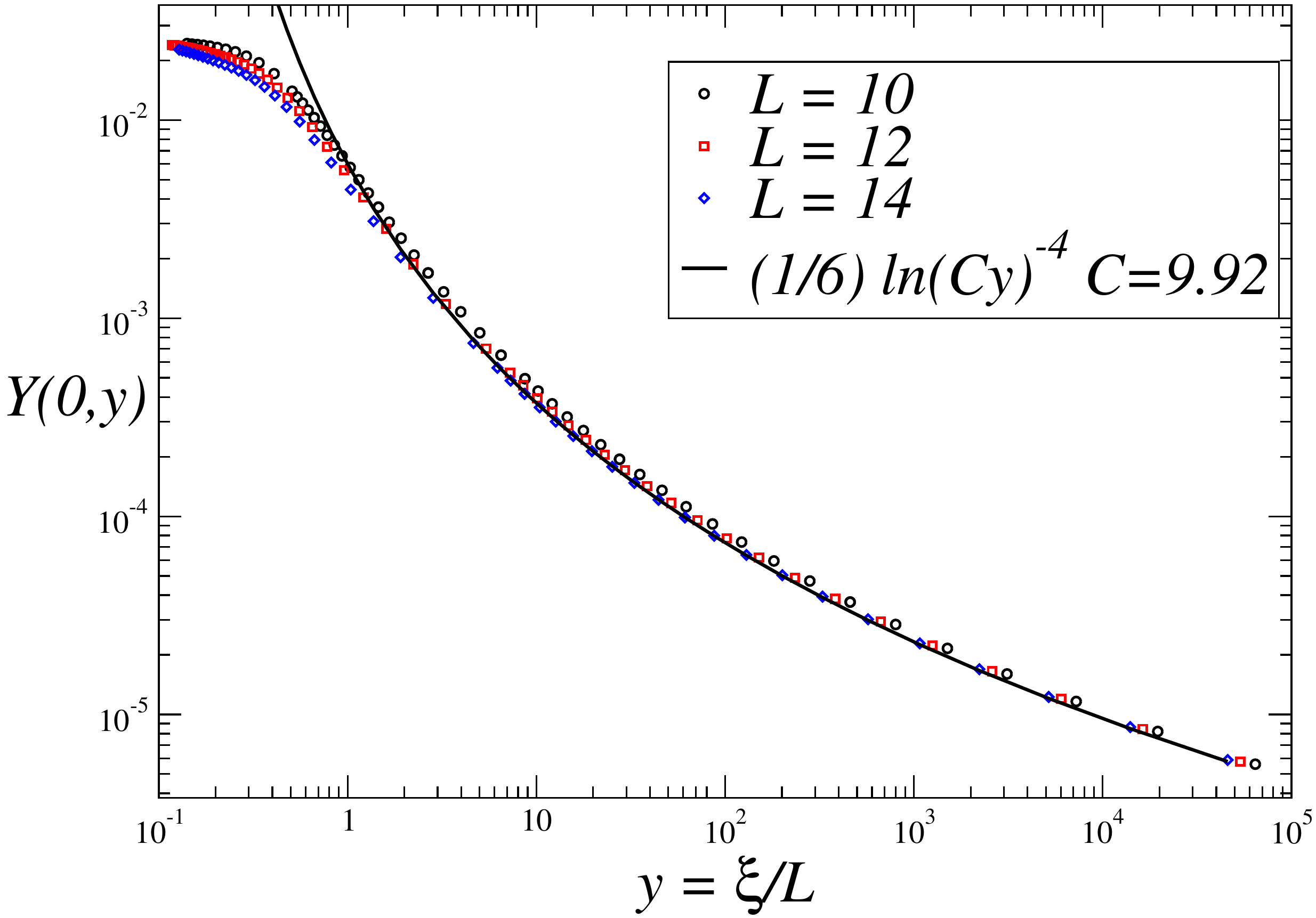} 

\caption{(color online). Finite-size scaling function $Y\left(0,y\right)$
as a function of the normalized correlation length $\xi/L$. The solid
line is obtained by requiring consistency with the non-interacting
value at $U=0.$ The constant $C$ is obtained by a best fit with
the numerical data (symbols). Boundary conditions are antiperiodic
when system size $L$ is a multiple of four, periodic otherwise. \label{fig:numerics}}

\end{figure}

\subsection{Numerical Analysis at half filling}

To study the scaling behavior of the metric at half-filling, we turned
to exact, Lanczos, diagonalization. We find that the metric obeys
the scaling form Eq.~\eqref{eq:g_hyper-scaling} with the scaling
function $Y\left(0,y\right)$ plotted in Fig.~\eqref{fig:numerics},
for values of its argument ranging over six order of magnitudes. 

After calculating the ground state $\Psi_{0}\left(U\right)$ of Hamiltonian
\eqref{eq:ham-hubbard} with the Lanczos algorithm, the intensive
fidelity metric $g$ is obtained from the fidelity $F(U,U+\delta U)=|\langle\Psi_{0}(U)|\Psi_{0}(U+\delta U)\rangle|$
using \begin{equation}
g_{L}=\frac{2}{L}\frac{1-F(U,U+\delta U)}{\delta U^{2}}\ ,\end{equation}
with $\delta U=10^{-3}$. The above equation is a good approximation
to the limit $\delta U\rightarrow0$ as long as $\delta U\ll1/\sqrt{Lg}$
which was confirmed to be the case in all simulations. The function
$Y\left(0,y\right)$ is then obtained via $Y=g_{L}\left(d\ln\xi/dU\right)^{-2}$.
In order to have as many data points as possible, we analyzed sizes
of length $L=4n+2$ with periodic boundary conditions (BC's), while
for length $L=4n$ antiperiodic boundary conditions were used. Choosing
such boundary conditions at half filling, assures that the ground
state is non-degenerate even at $U=0$ %
\footnote{Note also that both these classes of fermionic systems with different
boundary conditions, can be mapped onto an interacting systems of
hard-core bosons with periodic boundary conditions. This clarifies
further why it is possible to compare data obtained with different
boundary conditions.%
}. 

The behavior at large $y=\xi/L$ can be obtained by requiring consistency
with the value obtained in the free case $U=0$. Then the scaling
function must have the following limiting form\[
Y\left(0,y\right)=\frac{1}{6}\left[\ln\left(Cy\right)\right]^{-4},\quad y\to\infty,\]
where $C$ is a constant. Since when $y\to\infty$ $\ln\xi\sim2\pi/U$
the divergence in $\left(d\ln\xi/dU\right)^{2}$ cancels with the
logarithm above, and one obtains\begin{eqnarray*}
g\left(U\to0,L\right) & = & \frac{1}{24\pi^{2}}\left(\frac{\ln\left(\xi\right)}{\ln\left(C\xi/L\right)}\right)^{4}\\
 & = & \frac{1}{24\pi^{2}}.\end{eqnarray*}

Having computed the scaling function $Y\left(x,y\right)$ we could
ask what happens to the metric tensor as one approaches the BKT point
from the particular path $U\to0,\, n=1$. The smallest values of $y$
at our disposal are of the order of $y\sim10^{-1}$. Looking at Figure
\ref{fig:numerics}, on the basis of these data, it seems that  the
function $Y$ approaches a non-zero value $\lim_{y\to0}Y\left(0,y\right)$
as $y$ goes to zero. If this is the case, after taking $L\to\infty$
and $U$ small we would have\[
\lim_{L\to\infty}g\left(U,n=1,L\right)=\frac{4\pi^{2}}{U^{4}}Y\left(0,0\right).\]
Note however, that observing this divergence numerically can be very
hard as we must be in the region $L\gg\xi\left(U\right)$ which requires
huge sizes when the coupling $U$ is small.

\section{Conclusions}

In this article we analyzed the \emph{fidelity metric} in the zero-temperature
phase diagram of the 1D Hubbard model, with particular care at the
phase transition points. The fidelity metric quantifies the degree
of distinguishablity between a state and its neighbors in the space
of states, and as such it is expected to increase (or diverge) at
transition points. Special attention has been drawn to assess whether
the fidelity metric reveals signatures of the Mott-insulator transition
occurring at on site repulsion $U\to0$ and filling factor $n=1$.
Being a transition of infinite order, it is particularly difficult
to pin down since typical thermodynamic quantities are smooth (although
not analytic) at the transition. The point $U=0,\, n=1$ is particularly
singular in that it represents the limit of two completely different
physical regimes. On the line $U=0$ it is simply the half-filling
limit of a gapless free system, whereas fixing $n=1$ it represents
the limit of a complicated interacting massive system. Surprisingly,
we have shown that it is possible to interpolate between these two
regimes, and the fidelity metric defines a hyper-scaling function
which depends only on $x=\xi\left(U\right)$$\left(1-n\right)$. The
two regimes roughly correspond to $x\gg1$and $x\ll1$ respectively.

Away from half filling we have been able to compute the scaling function
integrating numerically Bethe-Ansatz equations, and we obtained analytic
expressions for the limits $x\to0,\infty$. The result implies that,
as a function of $U$ and $n$ separately, the fidelity metric has
no precise limit when $U\to0,\, n\to1$, but the scaling function
is well defined in term of the scaling variable $x$.

Precisely at half filling we computed the scaling function resorting
to exact diagonalization and upon introduction of another scaling
variable $y=\xi\left(U\right)/L$. With the numerical data at our
disposal, the scaling function appears to be smooth and non-zero around
$y=0$. As a a consequence, approaching the Mott point from the half-filling
line, the fidelity metric would display a singularity of the form
$U^{-4}$. A singularity of algebraic type has been observed also
in another instance of Kosterlitz-Thouless transition given by the
$XXZ$ model.

LCV would like to thank Cristian Degli Esposti Boschi for a critical
reading of the manuscript. The work of P.B.~has been partially carried
out in the framework of the PRIN project \emph{Microscopic description
of fermionic quantum devices}. 

\begin{appendix}

\section{\label{sub:asymptotic}}

Following Ref.~\citep{frahm90} the coefficient $a$ in Eq.~\eqref{eq:kappac}
is given by\[
a\left(U\right)=\frac{4\ln\left(2\right)}{Uf\left(U\right)}\]
 where \[
f\left(U\right)\equiv1-2\int_{0}^{\infty}\frac{J_{0}\left(x\right)}{1+e^{Ux/2}}dx,\]
and $J_{0}$ is a Bessel function. Using the following results \begin{eqnarray*}
\frac{1}{1+e^{\alpha x}} & = & \sum_{n=0}\left(-1\right)^{n}e^{-\left(n+1\right)\alpha x}\\
\int_{0}^{\infty}J_{0}\left(x\right)e^{-\beta x}dx & = & \frac{1}{\sqrt{1+\beta^{2}}},\quad\alpha,\beta>0,\end{eqnarray*}
we arrive at \[
f\left(U\right)=1+2\sum_{n=1}^{\infty}\left(-1\right)^{n}\frac{1}{\sqrt{1+n^{2}U^{2}/4}}.\]
Using the formalism of the Remnant Functions defined in Ref.~\citep{fisher72}
one realizes that $f\left(U\right)$ is related to $R_{1/2,0}^{\left(-\right)}\left(4U^{-2}\right)$.
With the help of the expansions in Ref.~\citep{fisher72} we obtain
the following expression

\[
f\left(U\right)=\frac{8}{U}\sum_{n=0}^{\infty}K_{0}\left(\frac{2\pi}{U}\left(2n+1\right)\right).\]
 It is now easy to obtain the desired expression, using the asymptotic
of the Bessel function\[
K_{0}\left(1/x\right)=e^{-1/x}\sqrt{\frac{\pi x}{2}}\left(1-\frac{1}{8}x+\frac{9}{128}x^{2}+O\left(x^{3}\right)\right).\]
Collecting the results together, we obtain at leading order \[
a\left(U\right)=\frac{\ln\left(2\right)}{\sqrt{U}}e^{2\pi/U}\left(1+O\left(U\right)\right).\]

\end{appendix}

\bibliographystyle{apsrev}
\bibliography{fid_hub}

\begin{thebibliography}{34}
\bibitem[{\citenamefont{Yang}(2007)}]{yang07}
\bibinfo{author}{\bibfnamefont{M.-F.} \bibnamefont{Yang}},
  \bibinfo{journal}{Phys.~Rev.~B} \textbf{\bibinfo{volume}{76}},
  \bibinfo{pages}{180403(R)} (\bibinfo{year}{2007}), \bibinfo{note}{note that
  recently the result of Yang for the fidelity metric has been corrected by a
  factor $1/2$ \cite{fjaerestad}}.
\end{thebibliography}

\end{document}